\documentclass[epj]{svjour}
\usepackage{graphics}
\usepackage{epsfig}
\usepackage{amssymb}
\begin{document}
\title{Scale-free avalanches in the multifractal random walk.}
\author{M. Bartolozzi\inst{1}
\thanks{\emph{Current affiliation:} Research Group, Grinham Managed Funds, Sydney NSW
2065, Australia}%
}                     
\offprints{marco.bartolozzi@gmf.com.au}          
\institute{Special Research Centre for the Subatomic
Structure of Matter (CSSM), University of Adelaide, Adelaide, SA 5005,
Australia }
%
%
\abstract{ Avalanches, or Avalanche-like, events are often
observed in the dynamical behaviour of many complex systems which
span from solar flaring to the Earth's crust dynamics and from
traffic flows to financial markets. Self-organized criticality
(SOC) is one of the most popular theories able to explain this
intermittent charge/discharge behaviour. Despite a large amount of
theoretical work, empirical tests for SOC are still in their
infancy. In the present paper we address the common problem of
revealing SOC from a simple time series without having much
information about the underlying system. As a working example we
use a modified version of the multifractal random walk originally
proposed as a model for the stock market dynamics. The study
reveals, despite the lack of the typical ingredients of SOC, an
avalanche-like dynamics similar to that of many physical systems.
While, on one hand, the results confirm the relevance of cascade
models in representing turbulent-like phenomena, on the other,
they also raise the question about the current state of
reliability of SOC inference from time series analysis.}
\PACS{
      {}{Self-organized criticality}   \and
      {}{Multifractal Random Walk}   \and
      {}{Wavelets} \and
      {}{Time series analysis} \and
      {}{Econophysics}
} 
\maketitle
\section{Introduction}


The theory of {\em self-organized criticality} (SOC) has been
developed in the late eighty's by Bak, Tang and
Wiesenfeld~\cite{Bak8788},  in order to explain the ubiquity of
power laws in nature. The key concept of SOC is that complex
systems ``naturally'' self-organize to a globally stationary
intermittent state in which avalanche-like events are power law
distributed.
 These features are similar to those found in
physical systems  at the critical point~\cite{Jensen}.
  The prototypical model of a system exhibiting SOC behaviour  is the
2D sandpile~\cite{Bak8788}. Here the cells of a grid are filled by
randomly dropping grains of sand (external driving).
  When the gradient
between two  adjacent cells exceeds a certain threshold a
redistribution of the sand occurs, leading to more instabilities and
further redistributions.  The characteristics of this system, indeed of all
systems exhibiting SOC, is  that the distribution of the avalanche
sizes, their duration and the energy released, obey power laws.

Remarkably, this kind of scale-free intermittent evolution is
similar to that observed in many physical and social systems.
Examples include astrophysical and geophysical
plasmas~\cite{flares,spplasma}, earthquakes~\cite{earthquake},
evolutions of species~\cite{Bak93}, traffic
dynamics~\cite{traffic}, wars~\cite{Roberts98} and  the stock
market~\cite{Turcotte99,Bak93b,Bak97,Feigenbaum03}  (a recent
review on the subject can be found in Ref.~\cite{Turcotte99}).
Despite the theoretical interest, reliable tests to prove the
presence of SOC in real systems  are still in their infancy.  Some
attempts have been made in the  contest of solar
flaring~\cite{Boffetta99}, astrophysical~\cite{Bruno01,Lepreti04}
and laboratory~\cite{Spada01,Antoni01}  plasmas and the stock
market~\cite{Bartolozzi05}.  These works, while leaving open the
question of a SOC behaviour, clearly show that the evolution of
these systems can be well described by an avalanche-like dynamics
characterized by  power laws in the avalanche size, duration and
waiting time between them. The presence of correlation between
laminar times, that is the time elapsed between two avalanches, in
particular,
 has raised objections to the
relevance of SOC in these contexts.  In fact,  due to the lack of
memory in the random external driving commonly  used in the
simulations of conservative SOC systems, the probability
distribution function (PDF) of laminar times actually follows an
exponential decay~\cite{Wheatland98}. However, for
non-conservative systems, power laws can still be observed in the
presence of temporal correlations of the avalanches near the SOC
state~\cite{Freeman00,Carvalho00}. Such temporal correlation could
also be due to the intrinsic memory process (possibly chaotic) in
the driver~\cite{DeLosRios97,Sanchez02}.

 Motivated by recent observations of avalanche-like
dynamics in financial time series~\cite{Bartolozzi05}, we
investigate a possible similar behaviour in the popular {\em
multifractal random walk} (MRW) originally proposed in
Ref.~\cite{Bacry01} as a paradigm for the stock market behaviour.
This model, although not presenting the characteristic mechanisms
of  SOC, such as a threshold triggering for the avalanches, is
able to reproduce most of the stylized features of the stock
market.
Moreover, the MRW belongs to a family of cascade-like models
widely used to reproduce the statistical features of the velocity
fluctuations in hydrodynamic turbulence and, therefore, the
discussions outlined in the next sections go beyond their
application to finance but can be extended to every complex system
which displaying a turbulent-like dynamics.

In the next section we introduce the asymmetric MRW proposed by
Chen, Jayaprakash and Yuan~\cite{Chen05}, which avalanche dynamics
is investigate in detail in the rest of the paper. In
Sec.~\ref{sec:wavelet} we introduce the method of analysis while
the results are exposed in Sec.~\ref{sec_analysis}. Discussions
and conclusions are left for the last section.

\section{The MRW Model: the CJY Version}
\label{sec_model}

Recently, the study of the stock market, seen as a  complex
system, has attracted the attention of many physicists (for
reviews see
Refs.~\cite{Mantegna99,Bouchaud99,Paul99,Feigenbaum03}). Its
dynamical behaviour is characterized by ``stylized facts" mainly
concerning the {\em logarithmic returns}, $r(t)= \ln \left[
P(t)/P(t-1) \right]$ (where $P(t)$ is the stock price), and their
absolute values, that can be regarded as a measure of the
 {\em volatility}, $v(t)=|r(t)|$.  Such stylized
facts have been used as a guide for validating phenomenological
models of stock price fluctuations. Among them,
 the appearance of ``fat tails" in the PDF of the logarithmic
 returns, related to frequent large fluctuations in price,
 and the long time correlations present in the
 volatility, a phenomenon  known as
{\em volatility clustering}, have been extensively investigated in
the econophysics
literature~\cite{Kaizoji00,Krawiecki02,Takayasu9798,Bartolozzi}.
The evidence of leptokurtic distributions in financial time series
leads to immediate analogies with the longitudinal fluctuations of
turbulent flows where a similar dynamics is observed, although
differences have been pointed out as well~\cite{Mantegna99}.
Motivated by the aforementioned evidence, {\em cascade models},
originally developed to reproduce the characteristic features of
intermittency in hydrodynamic turbulence, have been applied, with
success, to reproduce some stylized facts of the stock market
dynamics\footnote{In hydrodynamic turbulence ``cascade" refers to
the flow of energy from the largest scales, where it is injected,
toward the smallest ones where it is finally dissipated. In the
market contest, instead, it is assumed that there exists a flow of
information among the different temporal scales adopted by the
traders. A further discussion on the subject is outside the scope
of this work but the interested reader can refer to the seminal
book of Frisch~\cite{Frisch} for a general review of cascade
models in turbulence and
Refs.~\cite{Ghashghaie96,Arneodo98,Braymann00,Calvet01,Lux01} for
applications to the stock market.}.

In this framework, one of the most popular models is
 the MRW originally proposed by Bacry, Delour and Muzy~\cite{Bacry01}.
  The version that we are going to use in the present investigation has been proposed by
Chen, Jayaprakash and Yuan~\cite{Chen05}, referred to as CJY in
the rest of the paper.
 In the CJY model the returns are expressed by
\begin{equation}
r(t)=\delta_t  \, z_t,
\label{main_eq}
\end{equation}
where $z_t$ is a Gaussian random variable with zero mean and
unitary standard deviation, while $\delta_t$ represents the
one-step volatility. The dynamics of the model, and therefore the
capability to reproduce the stylized
 facts of the stock market, is related to the dynamics of the variable $\delta_t$:
small variations lead to an intermittent behaviour in $r(t)$, similar
to the one observed in the financial markets.
 Specifically, we can write $\delta_t$ as
\begin{equation}
\delta_t=\delta_0 \, \gamma^{n(t)},
\label{model}
\end{equation}
where  $\delta_0$ is related to the  amplitude of the fluctuations
while $\gamma$ to their intermittency. The term $n(t)$, the core
of the model, is a bounded random walk with increments
\begin{equation}
\Delta n(t)  = \eta_t +\alpha \,\Psi(t) -\beta \, \bar{\eta}.
\label{increment}
\end{equation}
Here
\begin{equation}
\Psi(t) =  K_1 \,
\eta_{t}-K_{N_c+1}\,\eta_{t-N_c-1}+\sum_{i=1}^{N_c}
[K_{i+1}-K_{i}] \, \eta_{t-i}, \label{psi}
\end{equation}
%
%
%
with $\eta_{t}$ independent random variables, with average
$\bar{\eta}$, which assume the values +1 with probability $p$ and
$-1$ with probability $1-p$. In our simulations
$\bar{\eta}=2p-1<0$.
This term, that alone can reproduce volatility clustering, has
been found to be necessary in order to reproduce some scaling
properties of the conditional fluctuations observed in financial
data~\cite{Chen05}.
 The second term in the increment, $\Psi(t)$,
  has the ability to recover the long-time correlations
of the market volatility and, from now on, we will refer to it as
the {\em multifractal increment}\footnote{Eq.~(\ref{psi}), is
related to the original formulation of the MRW in
Ref.~\cite{Bacry01} where the logarithmic variance is expressed in
terms of a convolution between a memory kernel and a random
process as reported in Ref.~\cite{Sornette03,Sornette_book}.}. Its
strength coefficient, $\alpha$, is related to the degree of
intermittency of the time series and, therefore, to the time scale
of process. The kernel used in Eq.(\ref{psi}) for the convolution
is $K_i=1/\sqrt{i}$ and we fix the memory steps to $N_{c}=1000$ in
the simulations. The last term of Eq.~(\ref{increment}),  controls
the drift rate in $n(t)$, instead, with strength $\beta$, and,
therefore, adds more flexibility to the model.  A more detailed
discussion on the present model with $\alpha=\beta=0$ can be found
in Ref.~\cite{Chen03}.  The parameter $\gamma$ is fixed to $1.05$
in all the simulations and, following Ref.~\cite{Chen05}, it is
linked to $\alpha$ and $p$ according to:
\begin{equation}
\alpha  = \frac{\alpha_0}{\ln(\gamma)},
\label{alpha}
\end{equation}
\begin{equation}
p =\frac{1}{1+\gamma^{2}}.
\label{eq::probability}
\end{equation}
These particular choices are appropriate for a correct
reproduction of the observed statistical features of the data.
Note also that, for the previous choice of the parameter $\gamma$,
namely 1.05, the probability distribution of the random variable
in the multifractal increment, Eq.(\ref{eq::probability}), becomes
slightly asymmetric, $p=0.4756$, in contrast with the symmetry of
the original MRW~\cite{Bacry01}.
 As previously mentioned, we
 impose reflecting boundaries for $n(t)$ in order to prevent the
realization of extremely large or small (unrealistic) fluctuations
in the simulation of the market activity,
 namely $ 0 \le n(t) \le n_{max}$, with $n_{max}=\ln(30)/\ln(\gamma)$.
By doing that, $\delta_t$ in Eq.(\ref{main_eq}) is bounded between
$\delta_0$ and $30 \, \delta_0$.

\begin{figure}
\vspace{1cm} \centerline{\epsfig{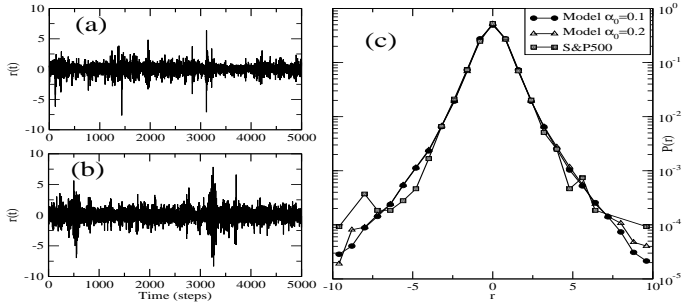}}
 \caption{(a) Part of the time series of logarithmic
returns for daily data of the S\&P500 index, used to calculate the
PDF in (c). The all set spans from 3/1/1950 to 18/7/2003. (b) Time
series generated from the model of Eqs.~(\ref{model})
and~(\ref{increment})
 with parameters $\gamma=1.05$, $\alpha_0=0.1$, $\beta=1.3$ and $\delta_0=1$.
The two time series shown in the plot have been standardized
according to $r(t) \rightarrow (r(t)-\langle r
\rangle)/\sigma(r)$, where $\langle \ldots \rangle$ and $\sigma$
represent, respectively, the average and the standard deviation
over the period in examination. (c) Comparison between the PDFs in (a)
and (b).
 The PDF generated with the parameters $\gamma=1.05$, $\alpha_0=0.2$, $\beta=4.0$
and $\delta_0=1$ is also shown. } \label{time_ser}
\end{figure}

In Fig.~\ref{time_ser} we compare the time series generated with
the model of Eqs.~(\ref{model}) and (\ref{increment}),
Fig.~\ref{time_ser} (b), and the time series of daily returns for
the S\&P500 index\footnote{The data have been collected from
3/1/1950 to 18/7/2003 for a total of 13468 samples. },
Fig.~\ref{time_ser} (a). The parameters used in the simulation,
and reported in the caption of the figure, have been chosen in
order to match the properties of the financial data set, as
underlined by the similarities in the PDFs of the two processes,
Fig.~\ref{time_ser} (c). Note, however, some discrepancies in the
tails of the PDFs. A possible explanation is that the dynamics of
the extreme events differs from the dynamics of the bulk of the
distribution (generated by the CYJ model, in this case) and they
could be interpreted as ``outliers"
\cite{Johansen98,Sornette_book}. However, finite size effect in
the relatively short time series of the S\&P500 should be also
considered.

\section{Wavelet Transform Filtering and Analysis Method}
\label{sec:wavelet}

As mentioned in the introduction, many complex systems show an
intermittent activity: quiescent periods are suddenly interrupted
by bursts of activity. This kind of non-stationary dynamics is
often related to multi-scale phenomena and most of the time
standard analysis techniques can fail to reveal some important
events that are localized in time or scale~\cite{Frisch}. This is,
for example, when using elementary filters: along with the noise
background also meaningful information can be filtered
out~\cite{Bartolozzi05}.

In order to overcome these problems, wavelet based techniques are
becoming more and more popular in complex systems
applications~\cite{Farge92}. This approach enables one to
decompose the signal in terms of scale and time units and so to
separate its coherent parts (or ``avalanches") -- that is, the
bursty periods related to the tails of the PDF -- from the
noise-like background, thus enabling an independent study of the
intermittent and the quiescent
intervals~\cite{Farge99,Spada01,Bruno01,Kovacs01,Bartolozzi05}.

%
%
The idea behind the wavelet transform is similar to that of
windowed Fourier analysis and it can be shown that the
scale parameter is indeed inversely proportional to
the classic Fourier frequency. The main
difference between the two techniques lies in the resolution in the
time-frequency domain. In  Fourier analysis the
resolution is scale independent, leading to aliasing of high and
low frequency components that do not fall into
the frequency range of the window. However,
in the wavelet decomposition the
resolution changes according to the scale (i.e. frequency).
At smaller scales the temporal resolution increases at the expense of frequency
localization, while for large scales we have the opposite. For this
reason the wavelet transform can be considered a sort of mathematical
``microscope''. While the Fourier analysis is still an appropriate method
for the study of harmonic signals, where the information is equally
distributed, the wavelet approach becomes fundamental when the
signal is intermittent and the information localized.

For a time series analysis it is often preferable to
use a discrete wavelet transform (DWT). The DWT can be seen as a
appropriate sub-sampling of the continuous wavelet transform (CWT)
by using dyadic scales.
That is, one chooses $\lambda=2^{j}$, for $j=0,...,L-1$,
where $L$ is the number of
scales involved, and the temporal coefficients
are separated by multiples of $\lambda$ for each dyadic scale, $t=n 2^{j}$,
with $n$ being the index of the coefficient at the $j$th scale.
The DWT coefficients, $W_{j,n}$, can then be expressed as
\begin{equation}
W_{j,n}=\langle f,\psi_{j,n}\rangle=2^{-j/2}\int f(u) \psi(2^{-j}u-n) du,
\label{dwt}
\end{equation}
where $\psi_{j,n}$ is the discretely scaled and shifted version of the
mother wavelet. The wavelet coefficients are a measure of the
correlation between the original signal, $f(t)$, and the mother
wavelet, $\psi(t)$  at scale $j$ and time $n$.
%
%
For the DWT, if the set of the mother wavelet and  its translated
and scaled copies form a complete orthonormal basis for all
functions having a finite squared modulus,  then the energy of the
starting signal is conserved in the wavelet coefficients. This
property  is, of course, extremely important when analyzing
physical time series~\cite{Kovacs01}. Following
Ref.~\cite{Bartolozzi05}, we use the Daubechies-4 as mother
wavelet~\cite{Daubechies88} for the analysis presented in the next
sections. Tests performed with different sets do not lead to any
qualitative difference in the results.
 A more comprehensive
discussions on the general properties of wavelets and their
applications are given in Refs.~\cite{Daubechies88,Farge92}.

The importance of the wavelet transform in the study of
turbulent-like signals lies in the fact that the large amplitude
wavelet coefficients are related to the extreme events in the
tails of the PDF, while the laminar or quiescent periods are
related to the  ones with smaller amplitude~\cite{Kovacs01}.  In
this way it is possible to define a criterion whereby one can
filter the time series of the coefficients depending on the
specific needs.  In our case we adopt the method used in
Refs.~\cite{Kovacs01,Bruno01,Bartolozzi05} and originally proposed
by Katul et al.~\cite{Katul94}.  In this method wavelet
coefficients that exceed a fixed threshold are set to zero,
according to
\begin{equation}
\tilde{W}_{j,n}=\left \{  \begin{array}{ccc} W_{j,n} & {\rm if} &
 W^{2}_{j,n}<C\cdot\langle W^{2}_{j,n}\rangle_{n},\\ 0 & {\rm
 otherwise},
\end{array} \right.
\end{equation}
here $\langle \ldots \rangle_{n}$ denotes the average over the time
parameters at a certain scale and $C$ is the threshold coefficient.
 Once we have
filtered the wavelet coefficients $\tilde{W}_{j,n}$ we perform an
inverse wavelet transform, obtaining a {\em smoothed}  version of
the original time series.

\begin{figure}
\vspace{1cm}\centerline{\epsfig{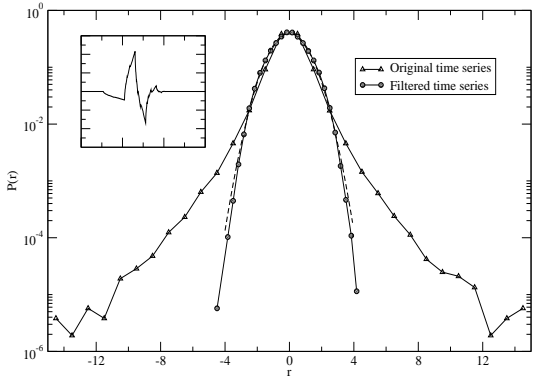}} \caption{PDFs of the original time series,
Fig.~\ref{ex_filtering} (a),
obtained with the model of Eq.~(\ref{model}) ($\gamma=1.05$,
$\alpha_0=0.1$, $\beta=1.3$
 and $\delta_0=1$) and its filtered version, Fig.~\ref{ex_filtering} (b).
A Gaussian is also plotted for
visual comparison (dashed line).
 The Daubechies-4 wavelet used for the analysis is shown in the inset. } \label{filter}
\end{figure}

The residuals of the original time series with the filtered one
correspond to the bursty periods, or avalanches,  which we aim to study.
An example of the filtering technique in terms of PDFs is given in Fig.~\ref{filter}.

\begin{figure}
\vspace{1cm}\centerline{\epsfig{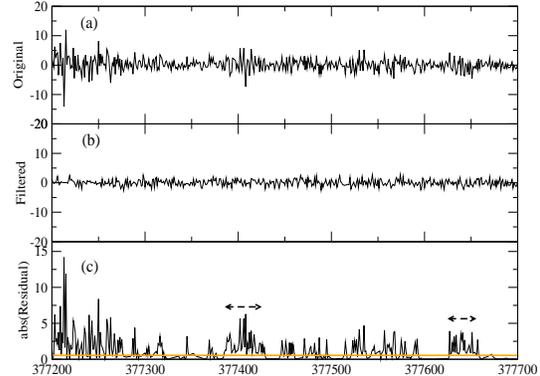}} \caption{ (a) Window of the time series obtained with
the model of Eq.~(\ref{model}) ($\gamma=1.05$, $\alpha_0=0.1$,
$\beta=1.3$
 and $\delta_0=1$). (b) Filtered version of the time series in (a)
 obtained with the Daubechies-4 wavelet.
In order to properly identify the avalanches we create a residual
time series by subtracting the filtered time series from the
original one. This noise removal technique presents advantages to
standard threshold methods when applied to multiscale
systems~\cite{Bartolozzi05}.
 (c) Absolute value of the residuals, or residual volatility, used to
extract the avalanches, as explained in the text. Two examples of
avalanches, among the many preset, are underlined by the dashed
lines in (c).
 The horizontal line slightly above zero represents the small
 threshold $\epsilon$. } \label{ex_filtering}
\end{figure}

 Once we have isolated the noise part from our signal
series we  are able to perform a reliable statistical analysis on
the {\em avalanches} of the residual time series. In particular,
we define the avalanches as the set of events during which the
volatility of the residual time series, $v_{res}(t)\equiv
|r_{res}(t)|$, is constantly above a positive small threshold,
$\epsilon \approx 0$. It is also possible to find an optimal value
for the choice of the threshold parameter $C$ in a way that the
filtered time series is, as close as possible, uncorrelated
Gaussian noise. From previous studies, this parameter is found to
be $C \sim 1$~\cite{Bartolozzi05}. In any case, the resulting
statistical analysis is qualitatively unchanged as long as $0
\lesssim C \lesssim 4$~\cite{Kovacs01,Bartolozzi05}. A graphic
example of the procedure for extracting the avalanches
 is illustrated in Fig.~\ref{ex_filtering}.

In analogy with the dissipated energy in a turbulent flow, we define the size
of an avalanche, $E$, as the integrated value of the squared volatility
over each coherent event of  the residual time series.  The duration,
$D$, is defined as the interval of time  between the beginning and
the end of a coherent event, while the laminar time, $L$, is the
time elapsing between the end of an event and the beginning of the
next one.

\section{Time Series Analysis}
\label{sec_analysis}

In the previous section we have seen how the wavelet multi-scale
filtering technique is an excellent tool to remove uncorrelated
Gaussian noise from an input signal.  In particular, this
filtering method becomes relevant whenever the examined time
series presents an intermittent behaviour, that is an irregular
switching, between periods characterized by large fluctuations and
noise-like ones. By using this technique, the avalanches, which
characterize an emergent behaviour in the dynamics, are
highlighted at the expense of the uninteresting background. In
this way we can make a proper statistical analysis of the
quantities that characterize these coherent events,
Fig.~\ref{ex_filtering}.

The statistical study of the avalanches identified with the
wavelet technique is of great interest, not only because this
would further test the capability of  this model to reproduce the
stylized facts of the stock market, but it could also shed some
light on the relevance of this test in distinguishing between SOC
and non-SOC processes in a time series analysis.
%
%

The analysis is carried out by studying how the statistics of the
avalanches change as we tune the parameters of the CJY model. The
time series generated with this algorithm have a length $N \sim 5
\cdot 10^5$ ($N=2^{19}$). Moreover, we investigate the specific
relevance of each term composing the increments of the variable
$n(t)$, Eq.(\ref{increment}). In order to do this, we
independently analyze the time series generated with different
expressions for the increments, $\Delta n(t)$: for each simulation
we consider a random walk for $n(t)$ with boundaries $n_{min}=0$
and $n_{max}=\ln(30)/\ln(\gamma)$.

We also fix  $C=1$ as the threshold coefficient for the wavelet
analysis. This particular value of $C$ is close to the
optimization value in the de-noise procedure. However a different
choice of this parameter would not change the qualitative results
of our analysis~\cite{Bartolozzi05}.

\subsection{The Role of Multifractal Increments}

We first investigate the avalanche dynamics
 generated by the multifractal increments, that is the part of the
 CJY model that is related to the original formulation of the MRW.
 In this case Eq.(\ref{increment}) reads as
%
%
\begin{equation}
\Delta n(t)=\alpha \, \Psi(t),
\label{eq::multifractal_inc}
\end{equation}
where $\Psi(t)$ is given by Eq.(\ref{psi}) and the strength
coefficient is expressed by $\alpha = \alpha_0/\ln(\gamma)$ with
$\gamma=1.05$.

\begin{figure}
\vspace{1cm}
\centerline{\epsfig{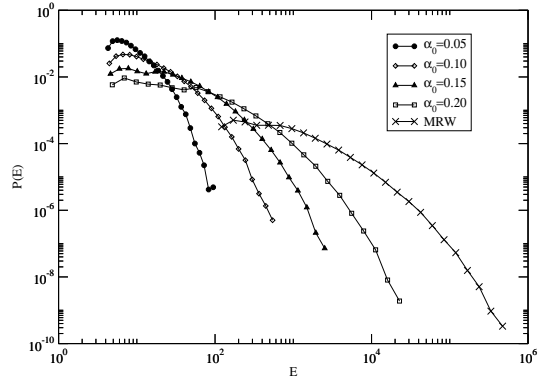}} \caption{PDFs of the avalanche sizes, $E$, as a
function of the parameter $\alpha_0$. For each value of this
parameter the time series of $E$ is noise-like and the
correspondent distribution displays an exponential decay. The
strength $\alpha_0$ controls the decay rate of the distribution.
The PDF for the MRW (X) with $\alpha=0.25$ is reported as well.}
\label{multif_en}
\end{figure}

\begin{figure}
\vspace{1cm}
\centerline{\epsfig{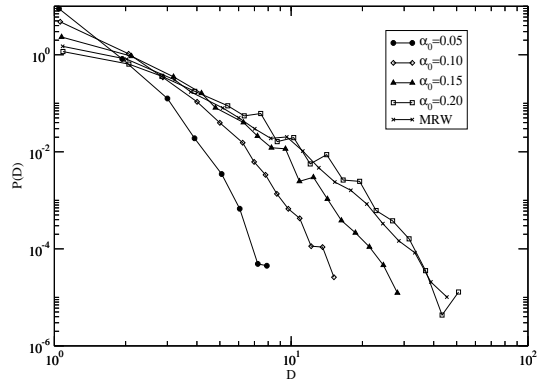}} \caption{PDFs of the duration, $D$, for the purely
multifractal model. No power law behaviour is observed and the
same considerations as for the size, $E$,
 hold in this case. The PDF
for the MRW (X) with $\alpha=0.25$ displays a similar behaviour. }
\label{multif_dur}
\end{figure}

\begin{figure}
\vspace{1cm}
\centerline{\epsfig{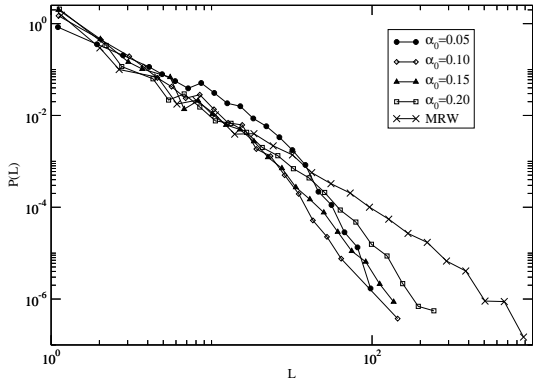}}
 \caption{Distribution of laminar times, $L$, in the
multifractal model. A slow convergence from an exponential decay
to a power law one is observed for $\alpha_0 \gtrsim 0.2$. The
exponent of the power law, in this case, is  $\nu \sim 2.3$. In
this case the PDF for the MRW (X), with $\alpha=0.25$, displays a
clearer power law behaviour compared to the CYJ version.}
\label{multif_lam}
\end{figure}

The avalanche analysis, resulting from the wavelet filtering, for
the size $E$, duration $D$, and laminar times $L$, is carried out
for different values of $\alpha_0$. This parameter, and $\alpha$
as a consequence, is related to the degree of intermittency of the
time series and, therefore, to the time scale of the process. For
$\alpha_0 \lesssim 0.05$, the dynamics of $r(t)$ is dominated by
noise and its PDF is a Gaussian. In the stock market contest, as
well as in turbulence, this corresponds to observe the
fluctuations at large scales. As we move this parameter toward
larger values, the time series of $r(t)$  becomes more and more
intermittent, giving rise to the large fluctuations which
characterize the broad tails of the PDFs of turbulent phenomena at
small scales. The results are shown in
Figs.~\ref{multif_en},~\ref{multif_dur} and ~\ref{multif_lam}.

From these plots it is possible to observe how the purely
multifractal model is not able to reproduce the power law
behaviour observed in real data sets. In particular, the PDFs for
$E$ and $D$ of the avalanches decay exponentially, independently
on the value of $\alpha_0$. This is an indication of the
randomness behind the avalanche generation process. The
distribution of laminar times in Fig.~\ref{multif_lam}, instead,
show a Poisson-like shape for small values of this parameter,
$\alpha_0=0.05$, while they start slowly to converge toward a
power law shape, $P(L) \sim L^{-\nu}$, for $\alpha_0 \gtrsim 0.2$.
The resulting exponent, $\nu \sim 2.3$, is similar to the one
found in the empirical studies~\cite{Bartolozzi05}.

For completeness, we study also the avalanche dynamics by using
the standard implementation of the MRW~\cite{Sornette03}. In this
case $\delta_t=\delta_0 e^{n(t)}$ and $\Delta n(t)$ is given by
Eq.~(\ref{eq::multifractal_inc}) where, this time, $\eta_t$ is a
Gaussian random variable with zero mean and unitary standard
deviation. The shapes of the PDFs of $E$ and $D$ resulting from
the analysis, shown in the Figs.~\ref{multif_en} and
~\ref{multif_dur} for $\alpha=0.25$,  display a similar $\alpha$
dependence as the CYJ version, as expected.
 However, the PDF of laminar
times, at least at small temporal scales, Fig.\ref{multif_lam},
display a clearer power low behaviour compared to the CYJ model.

\subsection{The Complete CJY Model}

\begin{figure}
\vspace{1cm}
\centerline{\epsfig{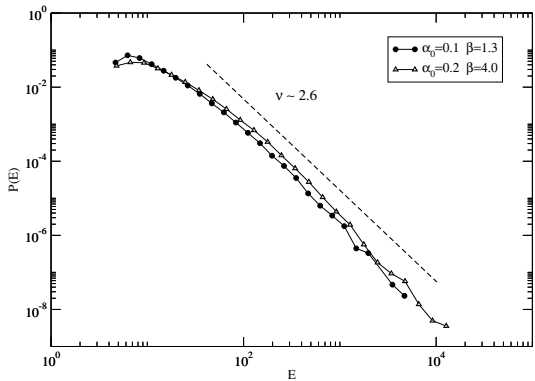}}
\caption{PDFs of the size, $E$, in the case $(\alpha_0=0.1, \beta=1.3)$ and
$(\alpha_0=0.2, \beta=4.0)$. A similar power law shape,
 with exponent $\nu \sim 2.6$, is observed for the two
distributions. A dashed line is also plotted for visual comparison.}
\label{comp_ene}
\end{figure}

\begin{figure}
\vspace{1cm} \centerline{\epsfig{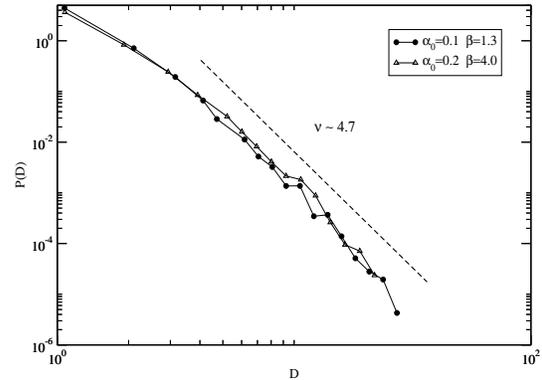}} \caption{PDFs for the avalanches duration, $D$, for
the same parameters of Fig.~\ref{comp_ene}. In this case a power
law with exponent $\nu \sim 4.7$, is observed.} \label{comp_dur}
\end{figure}

\begin{figure}
\vspace{1cm} \centerline{\epsfig{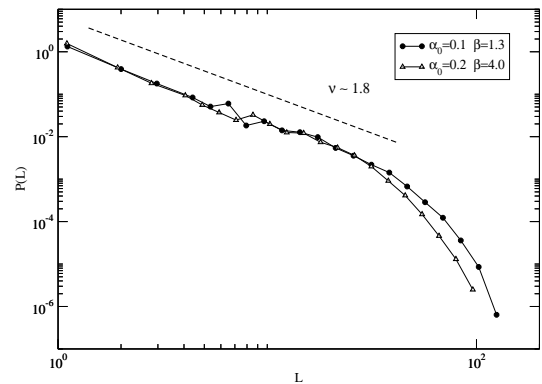}} \caption{PDFs for the laminar times between
avalanches, $L$. The parameters used are the same as in
Fig.~\ref{comp_ene}. A power law with exponent $\nu \sim1.8$ and a
cut-off at $L \sim 10^{2}$ is reported.} \label{comp_lam}
\end{figure}

We turn now our attention to the complete CYJ model of
Eq.(\ref{increment}). In this case, it has been
shown~\cite{Chen05} that a proper tuning of the parameters can
reproduce most of the stylized features of the stock market.  A
particular good agreement between the model and the empirical data
has been found by fixing $\gamma=1.05$ and $\delta_0=1.0$, for the
two couples of parameters $(\alpha_0=0.1, \beta=1.3)$ and
$(\alpha_0=0.2, \beta=4.0)$ as strength parameters for the memory
term and the drift respectively. It is, therefore, of particular
interest to explore the avalanche behaviour using these
parameters. The results of the analysis are shown in
Figs.~\ref{comp_ene}, \ref{comp_dur} and \ref{comp_lam} for $E$,
$D$ and $L$.

In this case, for some order of magnitude, we find a power law
behaviour for the quantities under consideration. This is in
qualitative agreement with the results found for the stock market.
In particular, the exponents of the power laws seem to be close to
the ones found for the analysis of the tick-by-tick Nasdaq E-mini
Futures~\cite{Bartolozzi05}.
Note that a scale-free avalanche dynamics has also been observed
in other reduced models of turbulence,  the {\em shell
models}~\cite{Boffetta99}, via a simple threshold technique.

This result can have important consequences regarding the possible
identification of SOC in the stock market, and other complex
systems in general, through a time series analysis.  In fact, we
have shown that an avalanche-like behaviour can also be observed
in models, such as the one presented in this work, in which the
characteristic ingredients of SOC, such as threshold dynamics, are
actually missing.   This, of course, does not rule out the
possibility of SOC but, nevertheless, more relevant and
discriminating tests become necessary.

\begin{figure}
\vspace{1cm}
\centerline{\epsfig{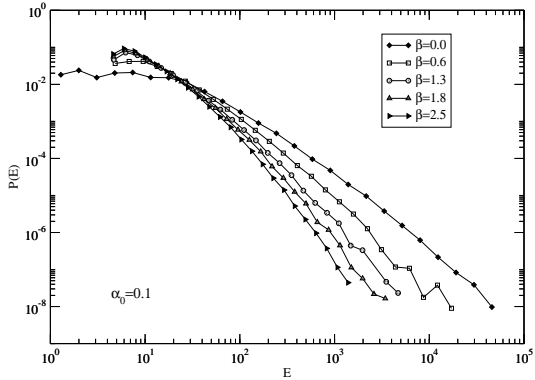}}
\caption{Dependence on the parameter $\beta$, with $\alpha_0=0.1$,
 for the PDF of the size, $E$, of the avalanche. The drift strength
controls the slope of the power law, which appear to go to saturation
for the higher values of $\beta$.}
\label{beta_ene}
\end{figure}

\begin{figure}
\vspace{1cm}
\centerline{\epsfig{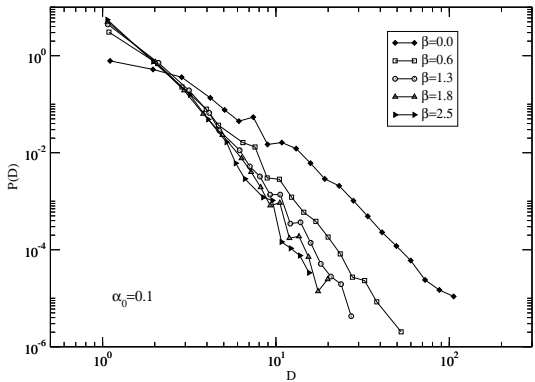}}
\caption{Dependence on the parameter $\beta$, with $\alpha_0=0.1$,
 for the PDF for the duration $D$ of the avalanches. Also in this case,
the drift strength controls the slope of the PDF and saturates for large $\beta$.}
\label{beta_dur}
\end{figure}

\begin{figure}
\vspace{1cm}
\centerline{\epsfig{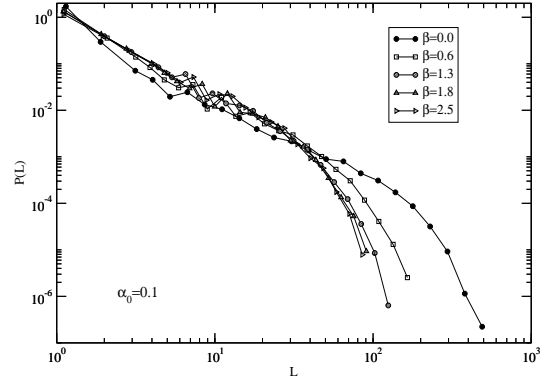}}
\caption{The exponent of the power law of the
laminar times between the avalanches, $L$, do not seem to be sensitive
to changes in the parameter $\beta$, which appear to be relevant just in changing
the cut-off of the distribution. }
\label{beta_lam}
\end{figure}

We further investigate our model by studying how the distribution of
 the $E$, $D$ and $L$ change with the drift strength $\beta$. In fact, different markets
 could have different power law exponents, no universality has been
 found until now, and we want to test the elasticity of the CYJ
 with respect to this parameter.  The PDFs  for $E$, $D$ and $L$ as
 functions of $\beta$ are reported, respectively, in
 Figs.~\ref{beta_ene}, \ref{beta_dur} and \ref{beta_lam}.

The analysis shows how the parameter $\beta$ plays an important
role in the dynamics of the model.  In fact, although the shape of
the PDFs for $E$ and $D$ are robust against variations of $\beta$,
the exponent changes with this parameter.  Higher values of
$\beta$ imply a larger value for $\nu$. The same arguments do not
hold for the statistics of the laminar times, $L$. In this case
the resulting distribution is pretty much independent of $\beta$
and could constitute a limit in the present formulation of the
model. A separate discussion is reserved for $\beta=0$.
In this case the filtering procedure, with $C=1$ as optimal value,
is not able to remove all the wavelet coefficients related to the
large fluctuations. As a consequence, the excess of kurtosis,
$K_e=\langle r^4 \rangle /\langle r^2 \rangle ^{2}-3$, of the
filtered time series, although still small in absolute value, $K_e
\sim 0.2$, becomes more than one order of magnitude larger that in
the previous analysis .
%
%
This means, to some extent, that there is not enough Gaussian
noise to be filtered out in the time series! Moreover, the shapes
of the PDFs related to the avalanche dynamics, show a behaviour
that is systematically different from the one observed once the
drift is included,   Figs. \ref{beta_ene}, \ref{comp_dur} and
\ref{beta_lam}.

\begin{figure}
\vspace{1cm}
\centerline{\epsfig{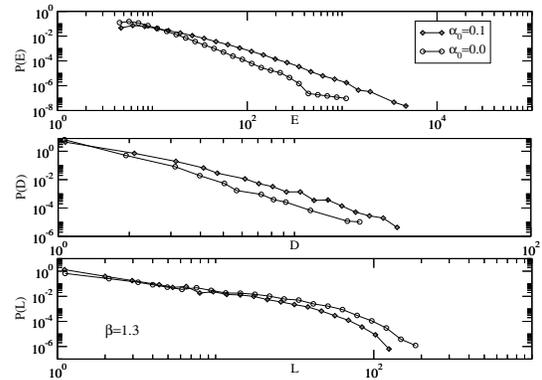}} \caption{ PDFs for $E$ (Top), $D$ (Middle) and $L$
(Bottom) for $\beta=1.3$ and $\alpha_0=0.1$, with multifractality,
and with $\alpha_0=0$, without the multifractal term. }
\label{no_memory}
\end{figure}

As a final note, we consider, for $\beta=1.3$ fixed, how our
analysis would change in the absence of the multifractal
increment, $\alpha \, \Psi(t)$, in Eq.(\ref{increment}). In doing
so, we compare
 the case with $\alpha_0=0.1$ and
$\alpha_0=0$. The results are shown in Fig.~\ref{no_memory}.

While the lack of the multifractal/memory term does not alter the
 distribution of laminar times between the clusters of volatility, it does
 increase the steepness  of the distribution of $E$ and $L$. This
 result is actually expected since this term builds the correlations
 inside periods of high volatility of the market time series, that is
 the so-called {\em volatility clustering}. By removing it we
 explicitly cut off part of the correlations inside the model,
resulting in shorter avalanches.

\section{Discussion and conclusion}

In the present work we investigated a possible avalanche-like
dynamics in an extended version of the popular  multifractal
random walk, the CJY,  proposed as a paradigm for the stock market
dynamics. We have been able to identify avalanche-like events in
the fluctuations generated by this model. Subsequently, the
statistical properties of these events have been estimated. The
identification of these clusters goes through an intermediate
passage where we use a wavelet filtering technique in order to
suppress the contribution of the noise background and, therefore,
enhance the precision of our measures.

The results show that, for a broad range of the parameters,
 the distribution of size, duration and
laminar time between avalanches follow a power law distribution.
 A very similar behaviour  has been found  in
empirical studies on financial time series~\cite{Bartolozzi05}.
Therefore, we confirm the relevance of the cascade models, and in
particular the CYJ version of the MRW, in modelling financial
market. Our results also extend beyond the financial environment
since this framework is quite general for describing dissipative,
intermittent, systems, such  as solar flaring or MHD turbulence,
where an avalanche-like dynamics has been observed as
well~\cite{Boffetta99,Bruno01,Lepreti04,Spada01,Antoni01}.

 Equally important, our results stress how models lacking of typical SOC
mechanisms can, nevertheless, manifest an avalanche-like
behaviour. In fact, the recognition of SOC ``patterns'' could be
an artifact of the identification method itself.
In this regard, it is worth mentioning that  simple threshold
techniques, when applied to time series generated by SOC-free
processes, can lead to the identification of avalanches with size
distributions (according to some specific definition) that are
power laws. For example, it is well known that the ``time for
first return to the origin"  of a random walk is power law
distributed~\cite{Sornette_book} despite the lack of any
correlation in the time series.
A more sophisticated example is reported in Ref.~\cite{Carbone04}
where it has been shown that the behaviour of a self-affine
fractional Brownian motion, when analyzed with a moving average
technique, can mimic the avalanche dynamics of the Dhar-Ramaswamy
sandpile. These results illustrate the technical ambiguity in the
identification of SOC from time series. As long as we do not have
any  {\em a priori} information about the underlying dynamics of
the system, it is very hard to tell if the avalanches that we
observe are a result of  a genuine SOC dynamics or any other
diffusive process.

In conclusion, SOC has been claimed, perhaps too loosely, to play
a role in many complex systems, but  there is no method that is
reliable enough to test its presence from the analysis of noisy
time series. This is a very relevant issue since, in practical
situations,  all the available information regarding a system is
encoded in its time series. Therefore, an extension of this
theoretical framework, which would enable the present gap with
empirical analysis
 to be filled, is of great practical importance and
would probably settle many speculations on the subject.

\section*{Acknowledgements}
The author would like to thank Kan Chen for his kind hospitality
at the National University of Singapore and for the helpful
discussions at the origin of this work. The author would like to
thank also Derek Leinweber, Dorin Ionescu and Louise Ord for a
careful reading of the manuscript. This work was supported by the
Australian Research Council.


\begin{thebibliography}{00}

\bibitem{Bak8788} P. Bak  C. Tang and K. Wiesenfeld,
 Phys. Rev. Lett. {\bf 59}, 381 (1987);
P. Bak,  C. Tang and K. Wiesenfeld, Phys. Rev. A {\bf 38}, 364 (1988).
\bibitem{Jensen}H. J. Jensen, \textit{Self-Organized Criticality:
Emergent Complex Behavior in Physical and Biological Systems},
(Cambridge University Press, Cambridge, 1998).
\bibitem{flares} E.T. Lu and R.J. Hamilton, Astrophys.J. {\bf 380},
L89 (1991); E.T. Lu {\em et al.}, Astrophys.J. {\bf 412},
841 (1993).
\bibitem{spplasma} T. Chang {\em et al.},
\textit{ Advances in Space Environmental Research, Vol.I},
(Kluwer Academic Publisher,AH Dordrecht, The Netherlands, 2003);
 A. Valdiva {\em et al.}, \textit{Advences in Space Environmental Research, Vol.I},
(Kluwer Academic Publisher,AH Dordrecht, The Netherlands, 2003).
\bibitem{earthquake}P. Bak and C. Tang, J. Geophys. Res. {\bf 94},
15 635 (1989); Sornette A. and Sornette D., Europhys. Lett. {\bf 9},
197 (1989); D. Sornette, P. Davy and A. Sornette, J. Geophys. Res. {\bf 95},
17 353 (1990); J. Huang {\em et al.}, Europhys. Lett. {\bf 41},
43 (1998).
\bibitem{Bak93}P. Bak and K. Snappen, Phys. Rev. Lett. {\bf 71}, 4083 (1993).
\bibitem{traffic} K. Nagel and H.J. Herrmann, Physica A {\bf 199}, 254 (1993);
 K. Nagel and M. Paczuski, Phys. Rev. E {\bf 51}, 2909 (1995);
T. Nagatani, J.Phys. A:Math.Gen. {\bf 28}, L119 (1995);
T. Nagatani, Fractals {\bf 4}, 279 (1996).
\bibitem{Roberts98} D.C. Roberts and D.L. Turcotte,
Fractals {\bf 6}, 351 (1998).
\bibitem{Turcotte99}D. L. Turcotte, Rep. Prog. Phys. {\bf 62},
1377 (1999).
\bibitem{Bak93b}P. Bak, K. Chen, J.Scheinkman and M. Woodford, Ric. Econ. {\bf 47}, 3 (1993).
\bibitem{Bak97}P. Bak, M. Paczuski and M. Shubik, Physica A {\bf 246}, 430 (1997).
\bibitem{Feigenbaum03} J. Feigenbaum, Rep. Prog. Phys. {\bf 66}, 1611 (2003).
\bibitem{Boffetta99} G. Boffetta {\em et al.}, Phys. Rev. Lett. {\bf 83}, 4662 (1999).
\bibitem{Bruno01} R. Bruno  {\em et. al.}, Planet. Space Sci.{\bf 49}, 045001 (2001).
\bibitem{Lepreti04} F. Lepreti  {\em et. al.}, Planet. Space Sci.{\bf 52}, 957 (2004).
\bibitem{Spada01} E. Spada {\em et. al.}, Phys. Rev. Lett. {\bf 86}, 3032 (2001).
\bibitem{Antoni01} V. Antoni  {\em et. al.}, Phys. Rev. Lett. {\bf 87}, 045001 (2001).
\bibitem{Bartolozzi05} M. Bartolozzi, D.B. Leinweber and A.W. Thomas, Physica A {\bf 350},
451 (2005); M. Bartolozzi, D.B. Leinweber and A.W. Thomas, Physica
A {\bf 370}, 132 (2006).
\bibitem{Wheatland98} M.S. Wheatland, P.A. Sturrock, J.M. McTiernan, Astrophys.J.{\bf 509}, 448 (1998).
\bibitem{Freeman00} M.P. Freeman, N.W. Watkins and D.J. Riley, Phys. Rev. E {\bf 62}, 8794 (2000).
\bibitem{Carvalho00} J.X. Carvalho and C.P.C. Prado, Phys. Rev. Lett.,
 {\bf 84}, 4006 (2000).
\bibitem{DeLosRios97} P. De Los Rios, A.  Valleriani and J.L. Vega, Phys. Rev. E {\bf 56}, 4876 (1997).
\bibitem{Sanchez02} R. Sanchez, D.E. Newman and B.A., Phys. Rev. Lett.
{\bf 88}, 068302-1 (2002).
\bibitem{Bacry01}E. Bacry, J. Delour and J.F. Muzy, Phys. Rev. E, {\bf 64}, 026103
(2001); E. Bacry, J. Delour and J.F. Muzy, Physica A, {\bf 299},
84 (2001).
\bibitem{Chen05}K. Chen, C. Jayaprakash and B. Yuan, arXiv:
physics/0503157.
\bibitem{Mantegna99}
R. N. Mantegna and H. E. Stanley, \textit{An Introduction to
Econophysics: Correlation and Complexity in Finance}, (Cambridge
University Press, Cambridge, 1999).
\bibitem{Bouchaud99}J.-P. Bouchaud and M. Potters, \textit{Theory of Financial Risk},
(Cambridge University Press, Cambridge, 1999).
\bibitem{Paul99} W. Paul and J. Baschnagel, \textit{Stochastic Processes: From Physics to Finance}, (Spriger-Verlag, Berlin,1999).
\bibitem{Kaizoji00} T. Kaizoji, Physica A {\bf 287}, 493 (2000).
\bibitem{Krawiecki02}  A. Krawiecki, J.A.  Holyst and D. and Helbing,
Phys. Rev. Lett.{\bf 89}, 158701 (2002); A. Krawiecki and J.A.
Holyst, Physica A {\bf 317}, 597 (2003).
\bibitem{Takayasu9798} H. Takayasu, A.-H. Sato and M. Takayasu, Phys. Rev. Lett. {\bf
79}, 966 (1997); H. Takayasu and M. Takayasu, Physica A {\bf 269},
24 (1999).
\bibitem{Bartolozzi} M. Bartolozzi and A.W. Thomas, Phys. Rev. E {\bf 69},
046112 (2004); M. Bartolozzi, D.B. Leinweber and A.W. Thomas,
Phys. Rev. E {\bf 72}, 046113 (2005).
\bibitem{Frisch} U. Frisch, \textit{Turbulence}, (Cambridge University Press, Cambridge, 1995).
\bibitem{Ghashghaie96} S. Ghashghaie {\em et al.}, Nature {\bf 381}, 767 (1996).
\bibitem{Arneodo98} A. Arneodo, J.F. Muzy and D. Sornette, Eur. Phys. J. B { \bf 2}, 277 (1998).
\bibitem{Braymann00} W. Breymann, S. Ghashghaie and P. Talkner, International Journal of Theoretical and applied Finance
 {\bf 3}, 357 (2000).
\bibitem{Calvet01} L. Calvet and A. Fisher, Journal of Econometrics {\bf 105}, 27
(2001); L. Calvet and A. Fisher, Review of Economics and
Statistics {\bf 84}, 381 (2002).
\bibitem{Lux01} T. Lux, Quant. Finance {\bf 1}, 632 (2001); T.
Lux, Int. J. Mod. Phys. C {\bf 15}, 481 (2004).
\bibitem{Sornette03} D. Sornette, Y. Malevergne and J.F. Muzy, Risk Magazine {\bf
16(2)}, 67 (2003).
\bibitem{Sornette_book} D. Sornette, \textit{Critical Phenomena in Natural Sciences, {\rm 2nd Edition}},
(Springer Series in Synergetics, Berlin, Germany, 2006).
\bibitem{Chen03} K. Chen, C. Jayaprakash, Physica A {\bf
324}, 207 (2003).
\bibitem{Johansen98} A. Johansen, D. Sornette, Eur. Phys. J. B { \bf 1}, 141 (1998).
\bibitem{Farge92} M. Farge, Annu. Rev. Fluid Mech. {\bf 24}, 395 (1992).
\bibitem{Farge99} M. Farge, K. Schneider and N. Kevlahan, Phys. Fluids {\bf 11}, 2187 (1999).
\bibitem{Kovacs01} P. Kov${\rm\acute{a}}$cs, V. Carbone and Z. V$\rm \ddot{o}$r$\rm \ddot{o}$s, Planet. Space Sci. {\bf 49}, 1219 (2001).
\bibitem{Daubechies88} I. Daubechies, Comm. Pure Appl. Math. {\bf 41} (7), 909 (1988).
\bibitem{Katul94} G.G. Katul {\em et al.}, \textit{Wavelets in Geophysics},
pp. 81-105, (Academic, San Diego, Calif. 1994).
\bibitem{Carbone04} A. Carbone, H.E. Stanley, Physica A {\bf 340}, 544 (2004).



\end{thebibliography}
\end{document}